\documentclass[11pt,twoside]{article}

\usepackage{asp2006}
\usepackage{epsf}
\usepackage{psfig}
\usepackage{lscape}

\begin{document}

\title{Asteroseismology and the Solar-Stellar Connection}

\author{Travis S.\ Metcalfe}

\affil{High Altitude Observatory, NCAR, P.O.\ Box 3000, Boulder, CO 80307}

\begin{abstract}
In March 2009, NASA will launch the Kepler satellite---a mission designed 
to discover habitable Earth-like planets around distant Sun-like stars. 
The method that Kepler will use to detect distant worlds will only reveal 
the size of the planet relative to the size of the host star, so part of 
the mission is devoted to characterizing other suns using 
asteroseismology. In this proceedings, I give a broad overview of the 
Kepler mission and the data that it will produce, with a special emphasis 
on how it could improve our understanding of solar and stellar dynamos. I 
conclude with an update on the development of a stellar modeling pipeline 
for interpreting asteroseismic observations.
\end{abstract}

\section{Song of the Stars}

In the region of Canada that is now Quebec and Ontario, a Native American 
tribe known as the Algonquin developed a detailed myth about the annual 
path of the Big Dipper around the north celestial pole, which they chanted 
in their ``Song of the Stars'':

\begin{quotation}
\noindent We are the stars which sing, \\
We sing with our light;  \\
We are the birds of fire, \\
We fly over the sky. 
\end{quotation}

\noindent If we think of asteroseismology as the study of ``stars which 
sing with their light'', then NASA's Kepler mission is poised to record 
the largest stellar choir ever assembled. In this proceedings, I will give 
a brief overview of the Kepler mission, followed by a review of the 
specific types of data the satellite is likely to produce which have the 
potential to improve our understanding of solar and stellar dynamos. I 
will conclude with an update on the work I have been doing to prepare for 
the massive wave of data soon expected from Kepler.

\section{Overview of the Kepler Mission}

If you visit NASA's homepage for Kepler ({\tt http://kepler.nasa.gov/}), 
the banner across the top of the page proclaims it to be ``a search for 
habitable planets.'' Indeed, the primary science goal of the Kepler 
mission is to find Earth-like planets with orbits inside the habitable 
zones of Sun-like stars. So why is asteroseismology a part of this mission 
at all? Well, Kepler aims to detect these habitable planets through high 
precision photometry of exoplanet transits. The small dip in the amount of 
light coming from the system is a measure of the size of the planet {\it 
relative} to the size of the star. Thus, for a high precision 
determination of the {\it absolute} size of the planet, the stellar radius 
needs to be measured reliably. Asteroseismology using the solar-like 
oscillations observed in these stars should be able to measure the 
absolute stellar radii with a precision of 2-3\% \citep{kbc08a}, leading 
to a comparable precision on the radii of any planets that are eventually 
discovered.

The Kepler mission is currently scheduled for launch in March 2009, and 
will target a single large field of view (105 square degrees) just above 
the Galactic plane in the constellation Cygnus. The idea is to look close 
enough to the plane that it can monitor $\sim$10$^5$ solar-type stars, 
which should be sufficient to discover dozens of Earth-like planets if 
they are at all common---but far enough out of the plane that it reduces 
the contamination from giants, which are less interesting from the 
standpoint of habitable planets. The satellite will stare at this one 
field for the entire nominal mission lifetime of 3.5 years---enough to 
detect three transits separated by 1 year in a reasonable sample of 
stars---but depending on the success of the program, it may optionally be 
extended for an additional 2 years. It will monitor the brightness of 
100,000 stars with sub-rastered images processed on-board into 30-minute 
sums, while a revolving selection of 512 stars can be monitored with a 
1-minute cadence to document solar-like oscillations. Initially, all of 
these short cadence targets will be specified by the Kepler Asteroseismic 
Science Consortium (KASC, {\tt http://kepler.asteroseismology.org/}), 
which is being organized through the University of Aarhus in Denmark and 
is open to anyone who signs NASA's non-disclosure agreement regarding 
possible exoplanet signatures discovered in the data. As the mission 
continues, some fraction of the short cadence targets will be reserved to 
characterize the host stars of exoplanets after they are discovered. A new 
selection can be specified every 3 months, so by the end of the mission 
there should be asteroseismic data on many hundreds of solar-type stars 
monitored continuously from a few months up to several years.

\section{New Data for Dynamo Modeling}

In this section I outline the various types of data that are likely to 
emerge from the Kepler satellite, with a special emphasis on information 
that could improve our understanding of solar and stellar dynamos.

\begin{figure}[t]
\plotone{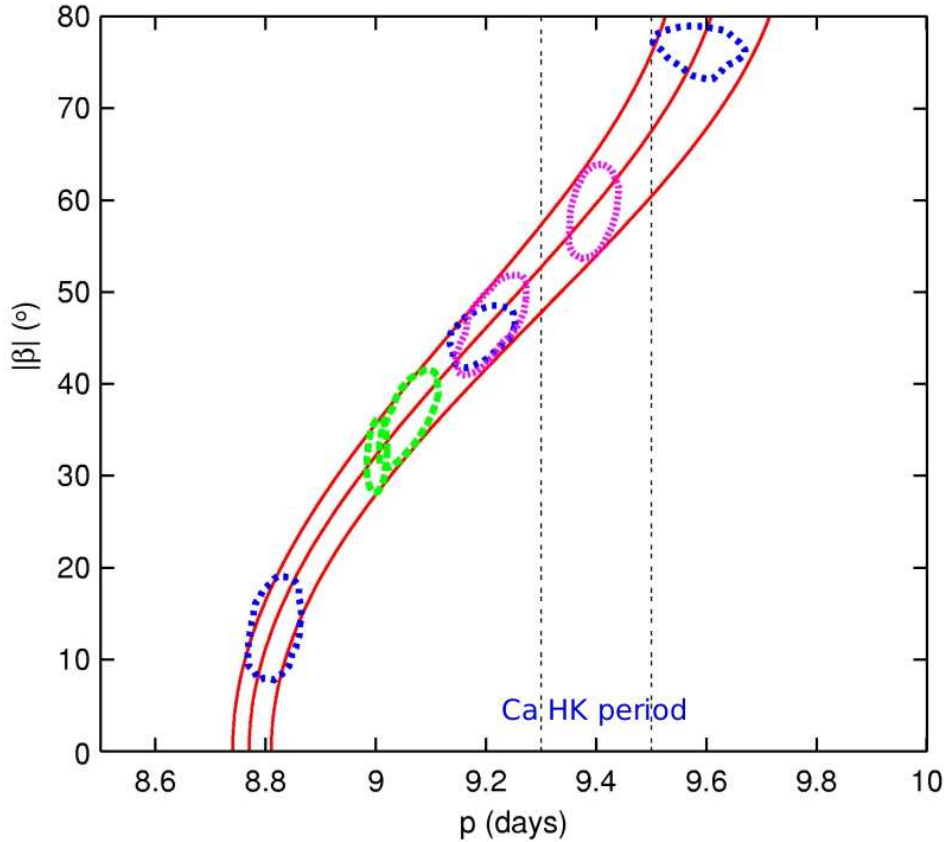}
\caption[Metcalfef1.eps]{The latitudes and rotation rates of star-spots on 
$\kappa^1$~Ceti from three seasons of MOST observations, showing the same 
pattern of surface differential rotation as the Sun \citep[adapted 
from][]{wal07}.\label{fig1}} 
\end{figure}

\subsection{Surface Differential Rotation}

Even without the short cadence data for asteroseismology, Kepler will 
yield high precision time-series photometry for many stars that will be 
sufficient to characterize the surface differential rotation through 
detailed spot modeling. Essentially, the photometry will be so precise 
that we will be able to see the signature of individual star-spots 
rotating into view---and the continuous monitoring will allow us to watch 
as spots at different stellar latitudes lap each other so we can derive 
their locations and rotation rates without ambiguity. A beautiful example 
of this technique comes from three seasons of photometry of the young 
solar-type star $\kappa^1$~Ceti from the MOST satellite \citep{wal07}. For 
each season of data, ranging from 15 to 30 days of observations, the MOST 
team fit the light curves with a stellar spot model. The rotation rates of 
the derived spots at various latitudes on the surface of the star exhibit 
the same functional form as the Sun's surface differential rotation, with 
spots near the equator moving around more quickly than spots closer to the 
pole (see Figure~\ref{fig1}). The Kepler data will allow similar 
differential rotation measurements for up to 100,000 solar-type stars, and 
over the lifetime of the mission this may even allow the construction of 
rudimentary ``butterfly diagrams'' showing the migration of activity belts 
through at least a fraction of the magnetic cycle.

\begin{figure}[t] 
\plotone{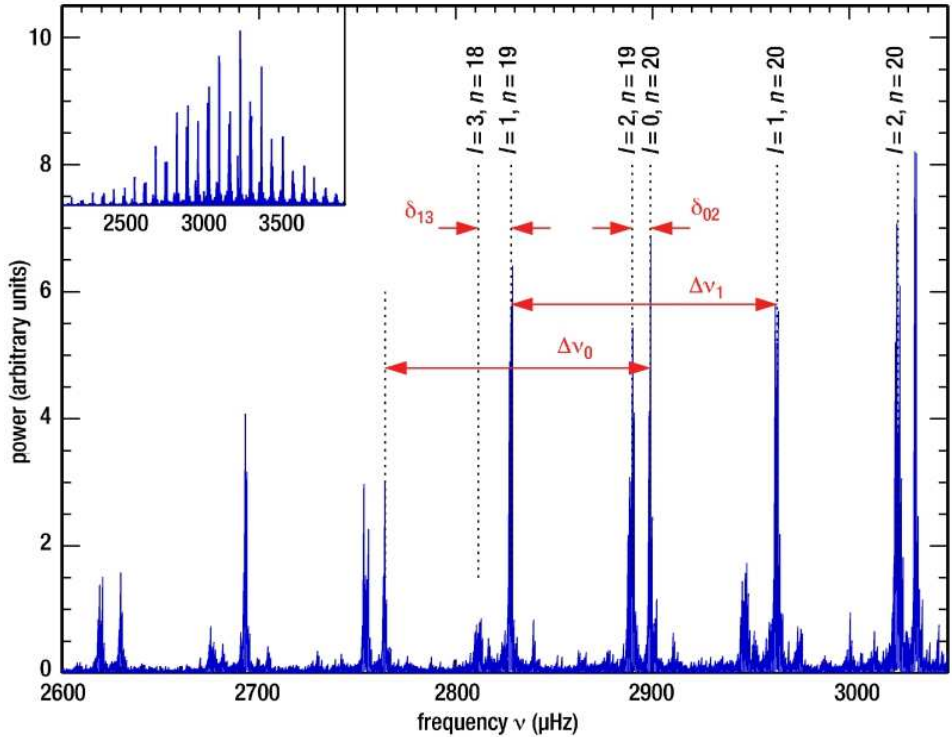} 
\caption[Metcalfef2.eps]{The pattern of oscillation frequencies seen in the 
Sun observed as a star, showing the characteristic large and small frequency 
spacings that indicate the mean density and age \citep[adapted 
from][]{et04}.\label{fig2}}
\end{figure} 

\subsection{Stellar Density and Age}

Among the stars that Kepler observes for asteroseismology, even those with 
the lowest signal-to-noise ratio---where individual oscillation 
frequencies may not be detected---should reveal the characteristic 
frequency spacings that probe the mean stellar density and age. 
Asteroseismic data of the Sun observed as a star, without spatial 
resolution across the disk, show a regular pattern of radial ($\ell=0$) 
and quadrupole ($\ell=2$) mode pairs interleaved with pairs of dipole 
($\ell=1$) and octupole ($\ell=3$) modes (see Figure~\ref{fig2}). The 
so-called large frequency separation---the spacing between consecutive 
radial overtones for modes with the same spherical degree---scales with 
the mean density of the star \citep{bg94}. When combined with non-seismic 
observables such as the effective temperature and luminosity, the large 
separation can lead to reasonable estimates of the stellar mass and 
radius. A radial mode travels through the center of the star, while the 
nearby quadrupole mode with a lower radial overtone does not. This defines 
the so-called small frequency separation, which is sensitive to interior 
chemical gradients near the center of the star caused by nuclear 
processing of hydrogen into helium. This is a good proxy for the stellar 
age, and will allow Kepler to measure ages for these stars with a 
precision of about 10\% of the main-sequence lifetime---compared to 
typical uncertainties of a factor of two for non-seismic age 
estimates. Together, these two characteristic spacings will significantly 
improve observational constraints on the evolution of magnetic activity 
and rotation across the entire region of the \mbox{H-R} diagram where 
solar-like oscillations are excited.

\begin{figure}[t] 
\plotone{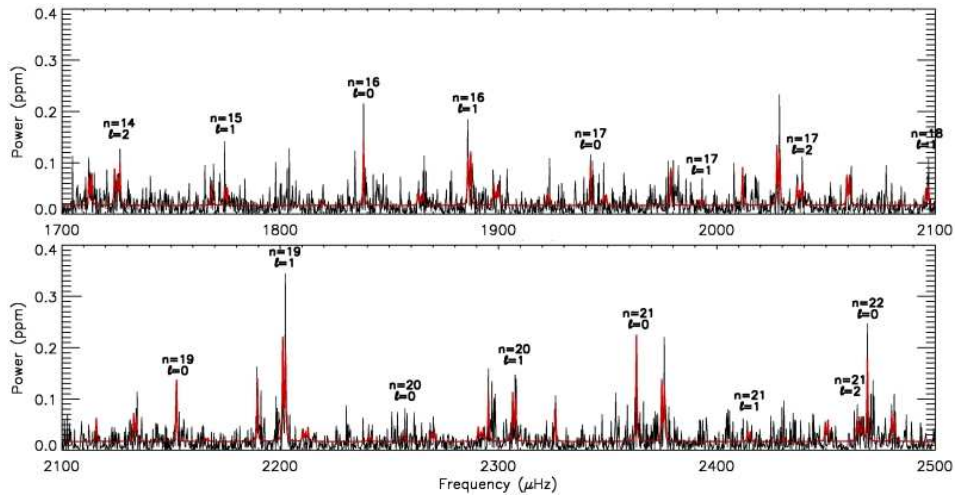} 
\caption[Metcalfef3.eps]{The rotational splitting as a function of radial 
overtone can provide indirect evidence of radial differential rotation 
since the modes sample slightly different regions of the star
\citep[adapted from][]{fle06}.\label{fig3}}
\end{figure} 

\subsection{Radial Differential Rotation}

For the brighter asteroseismic targets where the individual oscillation 
frequencies are detectable, the time series should be long enough to 
resolve rotational splitting of the modes into multiplets for stars with 
rotation rates between about 2 and 10 times the solar rate \citep{bal08}. 
Slower rotation makes it difficult to resolve the individual components of 
each multiplet from their strongly overlapping Lorentzian profiles, while 
faster rotation produces a splitting that is comparable to the small 
separation---creating some ambiguity in the mode identification. A 50-day 
photometric time series of $\alpha$~Cen~A from the WIRE satellite 
successfully resolved the rotational splitting in this star \citep[see 
Figure~\ref{fig3};][]{fle06}. Measurements of the rotational splitting as 
a function of radial overtone can indirectly probe radial differential 
rotation, since the various modes sample slightly different 
(but overlapping) regions of the star. More directly, even with the 
limited set of low-degree oscillation frequencies that are available for 
distant stars, it is possible to construct inversion kernels that might 
detect a rapidly rotating core \citep{gk93}, although more recent work 
suggests that a significant detection may require unrealistically strong 
differential rotation \citep{cha99}.

\begin{figure}[t] 
\plotone{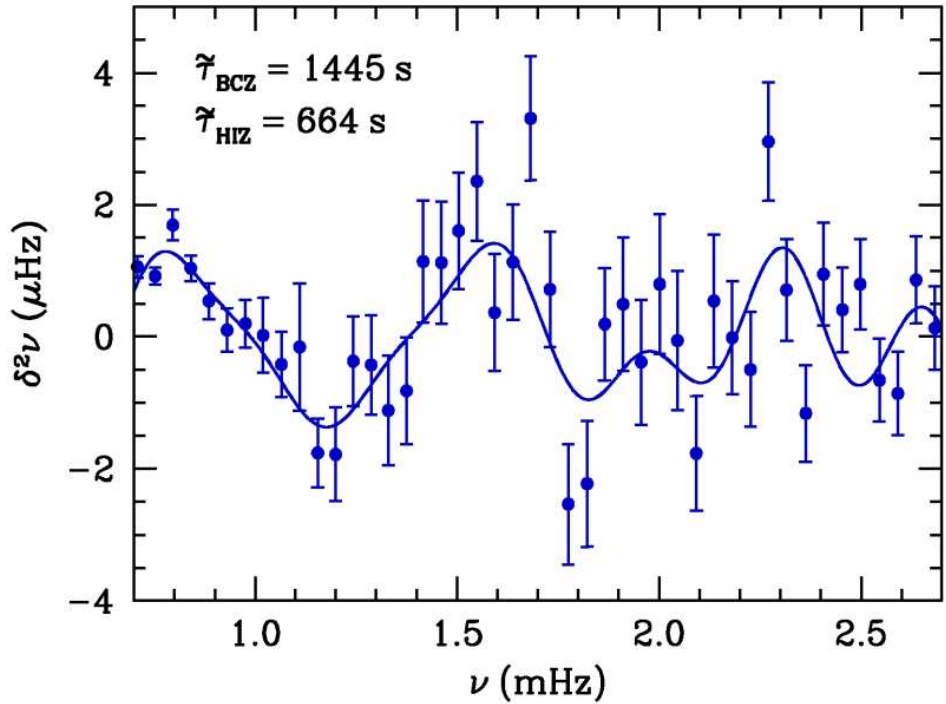} 
\caption[Metcalfef4.eps]{The base of the surface convection zone creates 
an oscillatory signal in the second differences that can indicate the 
acoustic radius of this sharp feature from detailed modeling 
\citep[adapted from][]{bag06}.\label{fig4}}
\end{figure} 

\subsection{Convection Zone Depth}

For the very best and brightest asteroseismic targets, it may be possible 
for Kepler to obtain a frequency precision sufficient to measure the depth 
of the surface convection zone. This concept was recently demonstrated 
with simulated data for a 5-month observation of the star HD~49933 using 
the CoRoT satellite \citep{bag06}. The second differences measure 
deviations from uniform spacing of a sequence of modes with the same 
degree. Sharp features inside the acoustic cavity, such as the base of the 
surface convection zone or the helium ionization region, create 
oscillatory signals in the second differences. Through detailed modeling, 
it is possible to derive the acoustic radii of these sharp features (see 
Figure~\ref{fig4}). Although results have recently been published for a 
short CoRoT run on HD~49933 \citep{app08} and a 5-month run on the same 
field has now been completed, it is not yet known whether the simulated 
data were a reasonable approximation of the actual observations. The 
Kepler mission has the potential to measure convection zone depths by 
applying this technique to dozens and possibly hundreds of solar-type 
stars.

\begin{figure}[t] 
\plotone{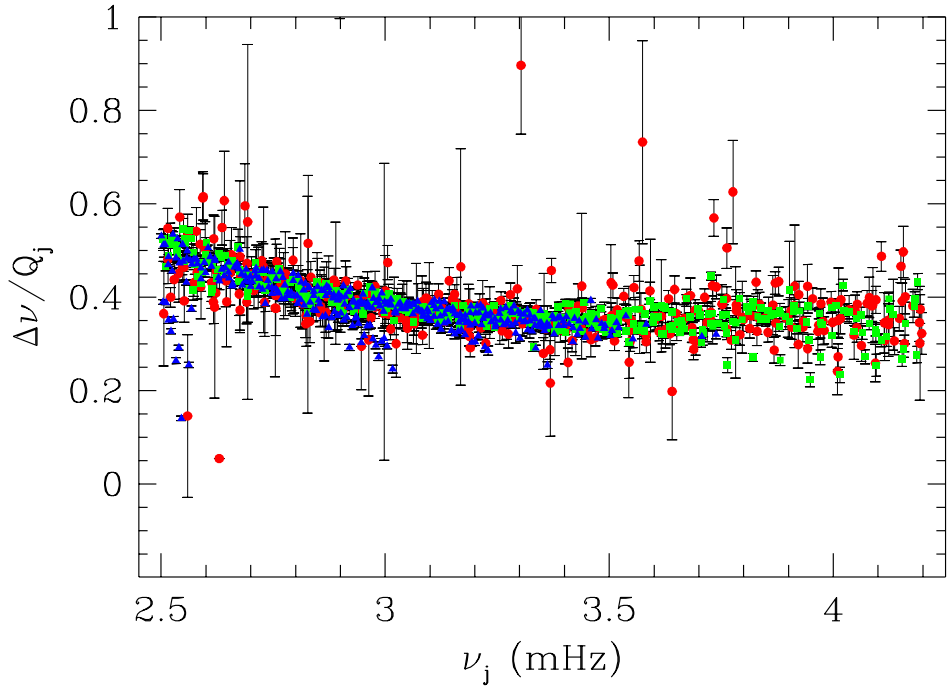} 
\caption[Metcalfef5.eps]{The solar cycle alters p-modes as a function of their 
frequency and spherical degree, but both dependencies largely disappear 
when normalized by an analytical model of the source \citep[adapted 
from][]{met07}.\label{fig5}}
\end{figure} 

\subsection{Cycle-induced Frequency Shifts}

The influence of the solar cycle on the Sun's p-mode frequencies was first 
detected nearly two decades ago \citep{lw90}, and was shown to depend on 
both the frequency and the spherical degree of the mode. Even the lowest 
degree p-mode frequencies from Sun-as-a-star observations show a strong 
correlation with the solar cycle \citep{sal04}, so there is good reason to 
believe that we will see similar frequency shifts in other stars. In 
principle, these shifts might provide unique constraints on the underlying 
physical mechanism. Recently, \cite{met07} developed a parameterization of 
the source of these frequency shifts in the Sun, and demonstrated that it 
could account for most of the dependence on both frequency and spherical 
degree (see Figure~\ref{fig5}). They applied this model to observations of 
the subgiant star $\beta$~Hydri and, although the precision of the 
asteroseismic data was not sufficient for a quantitative test, they showed 
that the appropriately scaled model could qualitatively account for the 
frequency shifts observed in this star. The Kepler mission will document 
similar shifts in hundreds of solar-type stars, allowing us to move beyond 
a simple scaling from the solar case and gradually leading to a broader 
context for our understanding of the solar dynamo.

\section{Stellar Modeling Pipeline}

\begin{figure}[!t] 
\plotone{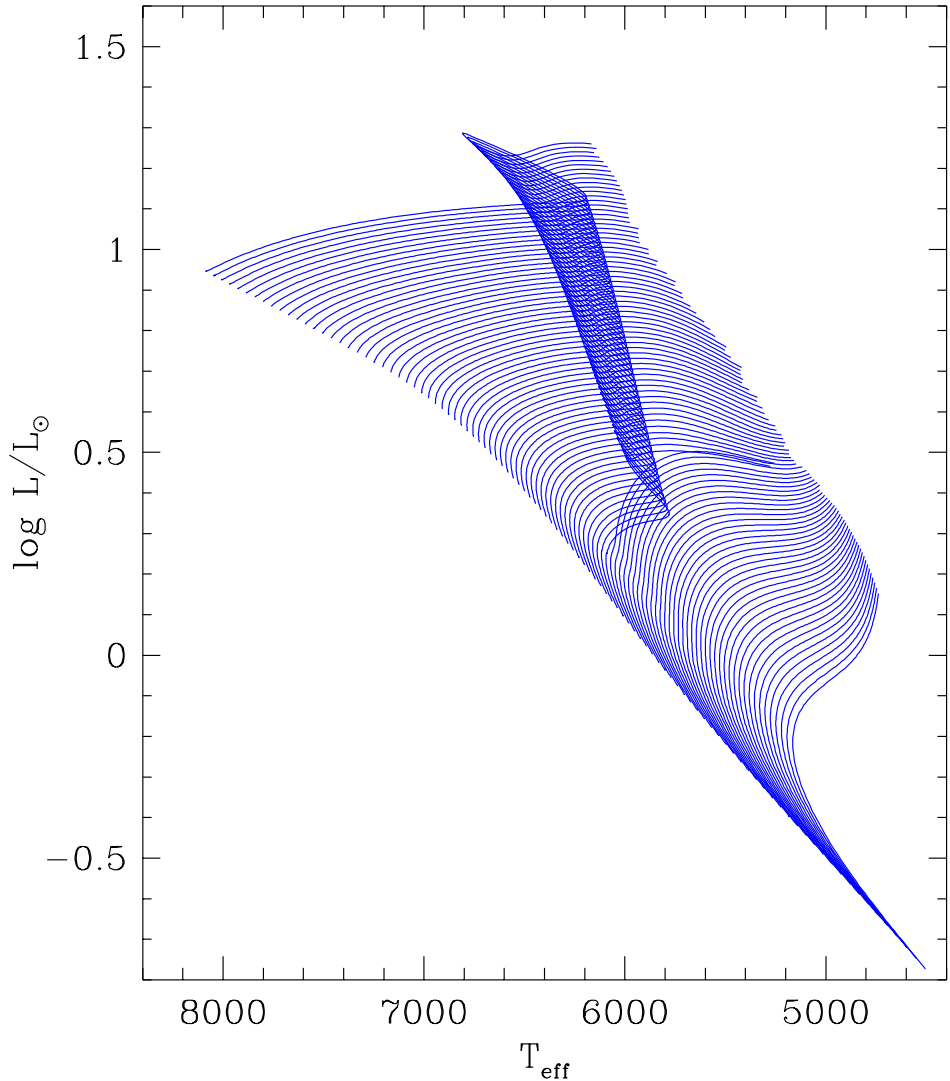}
\caption[Metcalfef6.eps]{Our stellar modeling pipeline explores a broad 
swath of the \mbox{H-R} diagram to find the globally optimal model for 
any set of asteroseismic data from solar-like oscillations, along with 
any other observational constraints.\label{fig6}}
\end{figure}

Over the past few years, I have been working with J{\o}rgen 
Christensen-Dalsgaard to develop a general purpose tool for modeling 
asteroseismic data. The basic idea is that, given an observed set of 
solar-like oscillation frequencies and non-seismic constraints, this 
software pipeline will return the optimal set of model parameters. It uses 
a genetic algorithm to explore a broad swath of the \mbox{H-R} diagram 
(see Figure~\ref{fig6}), probing masses between 0.75 and 1.75~M$_\odot$, 
initial metallicities between 0.002 and 0.05 sampled evenly in the log, 
initial helium mass fractions between 0.22 and 0.32, mixing-length 
parameters between 1 and 3, and evolutionary states from the zero age main 
sequence to the base of the red giant branch. By simultaneously matching 
all of the data, the method can find the optimal balance between 
asteroseismic and other constraints.

\subsection{Application to BiSON Data}

\begin{figure}[!t] 
\plotone{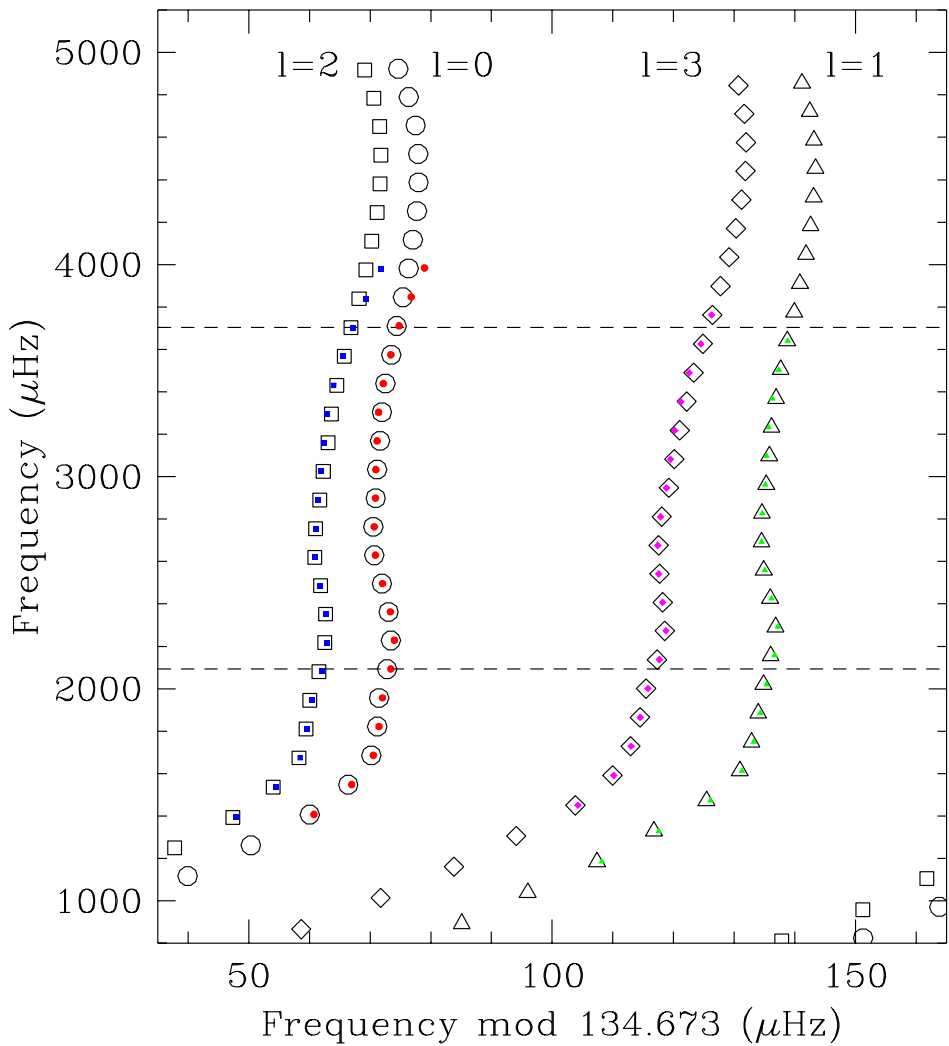}
\caption[Metcalfef7.eps]{An echelle diagram for the BiSON data (solid 
points) along with the optimal model from our asteroseismic modeling 
pipeline (open points), which only used the $\ell$=0-2 frequencies between 
the dashed lines for the fit.\label{fig7}}
\end{figure}

As a demonstration of the method, our first application has been to 
Sun-as-a-star observations from the BiSON network, using 36 oscillation 
modes with $\ell$=0-2 in a range of frequencies near 3 mHz and the solar 
luminosity and effective temperature with errors scaled up to what we 
expect from the Kepler mission. The genetic algorithm successfully matched 
the observed frequencies to better than 0.6~$\mu$Hz by including an 
empirical surface correction following \cite{kbc08b}, while simultaneously 
matching the temperature and age to $\pm$0.1\% and the luminosity and 
radius within $\pm$0.4\%. Although the fit used a limited range of 
frequencies and did not include $\ell$=3 modes, the optimal model also 
matched the modes with lower frequencies and higher degree (see 
Figure~\ref{fig7}) and recovered the known solar properties within 
reasonable tolerances.

\subsection{TeraGrid Portal}

Although the genetic algorithm requires far fewer model evaluations to 
find the global solution relative to a complete grid at the same sampling 
density, it is still quite computationally intensive---running on 512 
processors for about a week. To encourage a uniform analysis of the 
asteroseismic data from Kepler, we plan to make our pipeline available as 
a web-based tool tied to computational resources on the TeraGrid. This may 
help minimize any systematic errors that could arise if the data were 
analyzed by many people using different codes and optimization schemes. 
Through the web interface, users will be able to upload or enter their 
observational data, click a button, and a week later they will be emailed 
a file with all of the details of the optimal model. Since it requires 
about 100,000 model evaluations to find the best model for each star, we 
plan to build an archive that can be used for quick searches based on the 
accumulated incomplete grid. Over time, as we analyze many stars, this 
partial grid will grow more and more useful to identify reasonable 
models---particularly for new data sets that resemble earlier ones. 
Another service will simply be a user-friendly interface for the modeling 
code, allowing users to specify a single set of parameters and run one 
instance of the model, which will then be archived on the website. Of 
course, for users with access to large clusters or supercomputers, we will 
also provide the source code and documentation.

\section{Summary}

The Kepler mission promises to revolutionize the quality and quantity of 
asteroseismic data available for solar-type stars, and it will 
significantly contribute to research topics related to the solar-stellar 
connection. The mission needs asteroseismology to determine the absolute 
sizes of any potentially habitable Earth-like planets that might be 
discovered. It will also yield a wide variety of data that will be useful 
for calibrating dynamo models, sampling many different sets of physical 
conditions and evolutionary phases to complement the well-studied solar 
example. A uniform analysis of the asteroseismic data will help minimize 
systematic errors between the parameter values obtained for different 
stars, and such an analysis will be facilitated by our TeraGrid-based 
community modeling tool. The ``stars which sing with their light'' have 
many secrets to tell us, and the Kepler satellite will allow us to listen 
as never before.

\acknowledgements 
This work was initially supported by an NSF Astronomy \& Astrophysics 
Fellowship under award AST-0401441. Computer time for the stellar modeling 
pipeline has been provided by TeraGrid Development Allocation (DAC) 
AST-070031T, by NSF MRI grants CNS-0421498, CNS-0420873, CNS-0420985, the 
University of Colorado, and a grant from the IBM Shared University 
Research (SUR) program. The National Center for Atmospheric Research is a 
federally funded research and development center sponsored by the National 
Science Foundation.



\begin{thebibliography}{}

\bibitem[Appourchaux et al.(2008)]{app08} Appourchaux, T., et al.\ 2008, 
\aap, 488, 705

\bibitem[Baglin et al.(2006)]{bag06} Baglin, A., et al.\ 2006, ESA Special 
Publication, 624, 34

\bibitem[Ballot et al.(2008)]{bal08} Ballot, J., Appourchaux, T., Toutain, 
T., \& Guittet, M.\ 2008, \aap, 486, 867

\bibitem[Brown \& Gilliland(1994)]{bg94} Brown, T.~M., \& Gilliland, 
R.~L.\ 1994, \araa, 32, 37

\bibitem[Chaplin et al.(1999)]{cha99} Chaplin, W.~J., et al.\ 1999, 
\mnras, 308, 405

\bibitem[Elsworth \& Thompson(2004)]{et04} Elsworth, Y.~P., \& Thompson, 
M.~J.\ 2004, Astronomy and Geophysics, 45, 5.14

\bibitem[Fletcher et al.(2006)]{fle06} Fletcher, S.~T., et al.\ 2006, 
\mnras, 371, 935

\bibitem[Gough \& Kosovichev(1993)]{gk93} Gough, D.~O., \& Kosovichev, 
A.~G.\ 1993, IAU Colloq.~137: Inside the Stars, 40, 541

\bibitem[Kjeldsen et al.(2008a)]{kbc08a} Kjeldsen, H., Bedding, T.~R., \& 
Christensen-Dalsgaard, J.\ 2008a, IAU Symp.~253, in press 
(arXiv:0807.0508)

\bibitem[Kjeldsen et al.(2008b)]{kbc08b} Kjeldsen, H., Bedding, T.~R., \& 
Christensen-Dalsgaard, J.\ 2008b, \apjl, 683, L175

\bibitem[Libbrecht \& Woodard(1990)]{lw90} Libbrecht, K.~G., \& Woodard, 
M.~F.\ 1990, \nat, 345, 779

\bibitem[Metcalfe et al.(2007)]{met07} Metcalfe, T.~S., et al.\ 2007, 
\mnras, 379, L16

\bibitem[Salabert et al.(2004)]{sal04} Salabert, D., et al.\ 2004, \aap, 
413, 1135

\bibitem[Walker et al.(2007)]{wal07} Walker, G.~A.~H., et al.\ 2007, \apj, 
659, 1611

\end{thebibliography}
\end{document}